# LTE Cell Load Estimation Based on DCI Message Decoding

Daniil Chirkov, Artur Gaysin, Ivan Ashaev
Department of Radioelectronics and Telecommunications Systems
Kazan National Research Technical University named after A. N. Tupolev - KAI
Kazan, Russia
ChirkovDD@stud.kai.ru, AKGaysin@kai.ru, IPAshaev@kai.ru

*Abstract*— for modern mobile communication systems, the task of analyzing the parameters of quality of service and network information load remains one of the most popular. There are several ways getting this information in LTE networks. For example, by evaluation of test signals or estimation special reference signals in LTE frame. In this paper, it is proposed to obtain information about the cell load by decoding service messages of the DCI control information block. The main idea of this method is that the information load of LTE cell is estimated by decoding DCI messages in the LTE physical control channel and establishing the number of unique identifiers. For decoding the LTE signal, was used MATLAB LTE Toolbox and software-defined radio platform USRP 2920. In the work, the UE detection algorithm was tested on the signal generated using MATLAB software, and evaluated the information load of the network of a real LTE base station.

*Keywords*— LTE, DCI, USRP, Software-defined radio

## I. Introduction

Nowadays in telecommunications systems, improving the quality of service is one of the most important problem for research. This is particularly acute in the fourth - generation mobile networks-LTE, because here in one network there is transmitting of heterogeneous information. The main task of this project is estimation the information load of LTE cell. This will allow evaluating the quality of service more effectively.

There are different methods of estimation. A direct approach to solving this problem is to cooperate with the network organizers and gain access to the equipment and internal information of the operator. However, this is not always acceptable. Another method for assessing network congestion is to analyze radio resource management (RRC) messages in the common control channel (CCCH) [1].In addition there is method, based on analyzing indicators of receiving power RSRQ (Reference Signal Received Quality) and CQI(Channel Quality Indicator) [2], which correlated with cell load. Another way – generation probe signals and estimation of cell information load by the characteristics of this signal response. [3]

However, these methods are static; the information about network can be obtained only in finite moments. Therefore, another approach that lets to get information about information load of cell in live time.

According to this method, the information load of LTE cell is estimated by detecting distinctive user IDs and the number of resource elements allocated to them by searching for DCI (Downlink Control Information) messages in the PDCCH (Physical Downlink Control Channel) channel [4].

In LTE, to determine the resources that are intended for it each UE (User Equipment) has known identifiers C-RNTI (Cell Radio Network Temporary Identifiers). DCI messages are scrambling CRC bits by C-RNTI for a given UE. [5] Allocation of the DCI message in the PDCCH is also definite with helps C-RNTI. On the subscriber device, the CRC checksum of the DCI message is calculated using C-RNTI and, if the procedure is successful, the subscriber device determines that this message is intended for it.

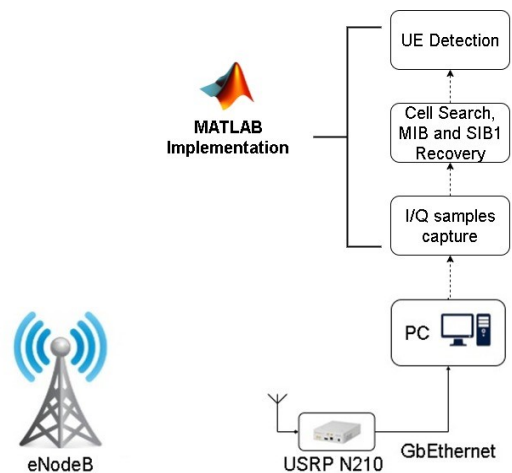

Fig. 1. Implementation of the DCI decoding method

The hardware implementation consists of transmission signals from the base station to the software-defined radio platform – USRP, which is connected via the GbEthernet interface to a computer with MATLAB software. Using MATLAB LTE Toolbox, the service information of the PDCCH are decoded in real time and information about the distribution and number of allocated resource blocks for each

UE are obtained by analyzing subscriber IDs. The Figure 1 is described the process of UE detection by decoding the DCI messages from the LTE base station.

The first sub goal is modeling and decoding signals using MATLAB LTE Toolbox. Next, signals received from the real LTE base station (eNodeB) will be decoded.

## II. UE DETECTION PROCEDURE ALGORITHM

The whole algorithm is divided into several parts: frequency detection and time synchronization, decoding of PBCH (Physical Broadcast Channel) and getting information from MIB (Master Information Block), system field allocation identification (PCFICH and PDCCH), DCI decoding from PDCCH and UE detection. All steps of this algorithm are depicted at the Figure 2. To decode DCI, algorithm should make all initial procedures of the typical UE. [4]

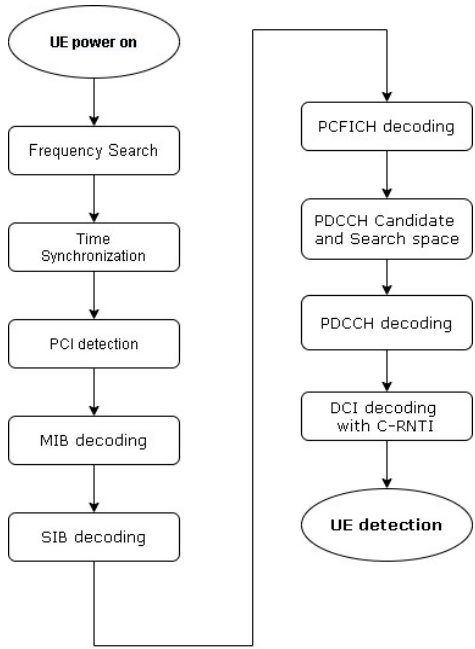

Fig. 2. Algorithm of UE detection by DCI decoding method

### A. Frequency detection and time syncronization

Firstly, then the equipment (UE) is started to initialization and authentication procedures with the base station, it must perform cell search and selection procedures and receive the main initial information of the system. [14]

At the beginning of this procedure, the mobile station finds the central frequency of the base station signal spectrum and synchronizes with the BS using the primary and secondary synchronization signals (PSS and SSS)[6]

*1) Primary synchronization signal (PSS)*

The PSS signal is needed for time synchronization, OFDM symbols, as well as for calculating the physical cell identifier (PCI). The PSS signal is transmitted in the 0 and 5-th subframes of each frame and for this transmission, the 62nd central subcarriers are used. PS PSS uses QPSK modulation. [7]

*2) Secondary synchronization signal (SSS)*

The SSS signal is also transmitted in the OFDM symbols of slots 0 and 5 on the central 62 subcarriers. [7] By receiving the SSS signal, the mobile station can determine the cell ID group-NID, which can take values from 0 to 167. After that, the mobile station calculates the cell ID (which is needed to determine the location of pilot signals) as follows [8]:

$$N_{ID(cell)} = 3 * N_{ID(1)} + N_{ID(2)} \qquad (1)$$

Therefore, there can be 504 different cell IDs. Receiving the SSS signal allows to achieve frame synchronization between the base station and the UE. This type of synchronization is achieved because different SSS sequences are transmitted in 0 and 5 subframes.

### B. MIB/PBCH decoding

A block of service information MIB (Master Information Block) is transmitted over the physical PBCH (Physical Broadcast Channel). MIB contains the value of the key parameters of the radio network. These parameters include channel width, PHICH channel configuration, and the System Frame Number (SFN). [8]

The channel width is specified in the number of resource blocks (which can be 6, 15, 25, 50, 75 or 100). After receiving this information, the mobile station (User Equipment, UE) can receive and decode the physical PDCCH and PCFICH channels. [9]

In the last MIB parameter, the most significant 8 bits of the SFN (System Frame Number) are passed. In general, an SFN can take values from 0 to 1023, so it needs 10 bits to transmit it. The last 2 bits of MS can be calculated after receiving four consecutive PBCH signals (for four frames, 40 ms). At the same time, to get all the other parameters from the MIB, it is enough to receive the PBCH channel only once in one frame (10 ms). The total size of a MIB block is 24 bits. After processing this data in the transmitter, 1920 bits are obtained from 24 bits. A block of 1920 bits is divided into 4 blocks of 480 bits each. Then each of the blocks is passed separately.

The physical PBCH channel is transmitted on 72 central subcarriers for 4 OFDM characters in the second slot of each frame. [9]

### C. PCFICH and PDCCH

*1) PCFICH (Physical Control Format Indicator Channel)*

This channel is contained in each subframe in the form of a Control Frame Indicator (CFI) and transmits the numbers of the OFDM symbols that are used to broadcast the messages of the PDCCH control channel. The CFI indicator contains 32 bits and is located in 16 resource elements of the first OFDM symbol of the descending frame. The signal is modulated by QPSK. The possible values for PCFICH - one, two, three, or four (rare case) symbols. [8]

*2) PDCCH (Physical Downlink Control Channel)*

The main task of Physical Downlink Control Channel (PDCCH) to carry the Downlink Control Information (DCI). The symbols are always at the start of each subframe. It Uses

QPSK modulation. Number of the symbols for PDCCH can be one, two, or three and it is depends from PCFICH. [8]

The REs (Resource Elements) allocated to PDCCH are grouped into group of 4 REs referred as quadruplets. RE quadruplets are grouped into CCE (Control Channel Elements). There are 9 quadruplets in one CCE. [10]

The number of CCE is depends on DCI format. The PDCCH format is selected according to the size of the DCI. DCI bits have a 16 bit CRC attached prior to rate 1/3 channel coding and rate matching. The PDCCH format must offer sufficient capacity to avoid puncturing the DCI bits too heavily during rate matching. [11]

### D. DCI (Downlink Control Information) and C-RNTI

In LTE, the Physical Downlink Control Channel (PDCCH) transfers control information of the system by Downlink Control Information (DCI) messages. [8] The DCI messages transmit the information about uplink or downlink scheduling from the eNodeB to destination UEs, in according to that the UEs can obtain the information about the resources location into the traffic channels Physical Downlink Shared Channel (PDSCH) and Physical Uplink Shared Channel (PUSCH). In addition to resource block allocation (RB), the DCI also contains data about modulation and encoding schemes, process numbers of the hybrid automatic Repeat Request (HARQ) algorithm, and power control commands for the uplink channels.

*1) DCI formats*

There are several different formats of DCI message used in LTE in PDCCH. The DCI format shows exactly what information and for what frame configuration is used and transmitted over the downlink to the PDCCH. [11] The table 1 illustrates the main DCI formats.

TABLE I. DCI FORMATS

| Format type | Application |
|---|---|
| Format 0 | Used for scheduling of uplink traffic channel (uplink grant); |
| Format 1 | Used for scheduling a PDSCH code word. Only a single transport block can be scheduled here using resource allocation type-0/type-1; |
| Format 1A | Used for scheduling a PDSCH code word. Only a single transport block can be scheduled here using resource allocation type2 (localized or distributed); |
| Format 2 | Used for scheduling of PDSCH in closed loop spatial multiplexing; |
| Format 2A | Used for scheduling of PDSCH in open loop spatial multiplexing. |
| Format 3 | Uplink transmit power control with 2 bit power adjustment; |

The DCI formats 0 and 3 are used for uplink and it is prefer not to use them during the call load estimation.

*2) C-RNTI*

The C-RNTI is temporary subscriber ID in the cell identifier (Cell RNTI). [12] It is used to transmit data when the mobile station is active and connected to the network (RRC Connected). This identifier is also used for scrambling data transmitted over traffic channels (PDSCH and PUSCH) channels. During the handover procedure, a new value of the C-RNTI identifier is assigned to the mobile station (since the values of this identifier are unique only within one cell).

### E. UE detection

In the casual LTE networks, the operation to get the DCI information and C-RNTI happens like this: the UE are search all possible PDCCH candidates to find, which has scrambled by it uncial C-RNTI and decodes only this resource blocks. [8]

However, in this work these identifications are unknown, so it is necessary to get them before. Assume that the transmission happens without error, i.e. assume CRC check sum is equal to zero. Then the last 16 decoded bits comprise the C-RNTI. [13] For decoding, the DCI message needs to be separated from the 16 C-RNTI bits. Next, compute new CRC bits and mask them with the C-RNTI. Then re-encode the initial DCI message by calculated C-RNTI. Finally compare new message with the original coded DCI.

From the all decoded DCI message, there is the distribution of resource blocks for each uncial C-RNTI. Also the number of active UEs and the number of RBs can be defined for them by this information. [4] The information load of the cell shows the ratio of the resource blocks involved to the maximum available total number of channel resources, that is, it shows how efficiently the available cell resource is used.

The information load of the cell can be calculated as:
$$I_{frame} = N_{ADLRB(f)} / N_{DLRB(f)} \qquad (2)$$
There $N_{ADLRB(f)}$ – number of assigned resource blokes in frame and $N_{DLRB(f)}$ is the total number of resource blocks for downlink channel.

Evaluation of the network load factor to draw a conclusion about how well the mobile network coverage is provided by the operator of this base station. For example, if the cell load is too low, it may be more appropriate not to take up free frequency resources, but to allocate them for the operator of another network. At the same time, overloading or heavy congestion of the cell may indicate that the available frequency resources are insufficient and the current quality of service indicators may be greatly reduced.

### III. MATLAB SIMULATION OF UE DETECTION

To evaluate and test the performance of the method, it is necessary to test it with a known distribution of resource blocks and C-RNTI. Figure 3 illustrates the algorithm of modeling.

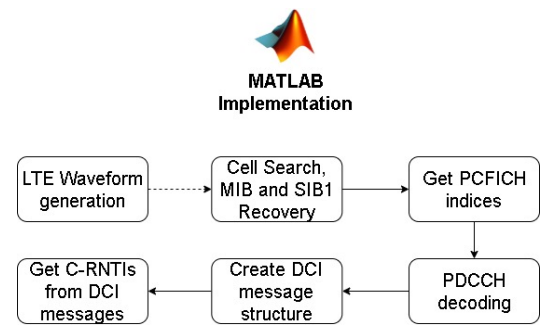

Fig. 3. MATLAB Simulation

## A. Simulation without noises

To do this, use MATLAB LTE Toolbox to create an LTE signal. The parameters of the signal:
- Number of RBs = 50;
- C-RNTI = 21;
- PCI = 27;

Its only downlink channel model. The frame structure is depicted at the Figure 4

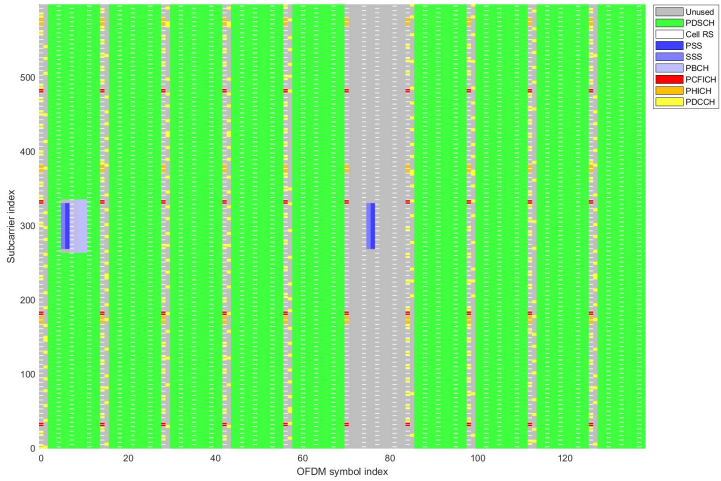

Fig. 4. Frame stricture of the modeling signal

Obtained results after UE detection, are shown in Table 2. In addition, the resource block allocation is depicted at the Figure 5. In the table, SFN is assigned to the format [frame.subframe] number. As can be seen from the table, without noises, there are zero errors and C-RNTI is decoded properly. Aslo, each subframe the RB are allocated at all 50 valuable positions.

TABLE II. DECODED DCI MESSAGE

| SFN | RNTI | Num of errors | DCI Format | Link Direction | RB Set |
|---|---|---|---|---|---|
| 0 | 0021 | 0 | Format1 | Downlink | [0...49] |
| 0.1 | 0021 | 0 | Format1 | Downlink | [0...49] |
| 0.2 | 0021 | 0 | Format1 | Downlink | [0...49] |
| 0.3 | 0021 | 0 | Format1 | Downlink | [0...49] |
| 0.4 | 0021 | 0 | Format1 | Downlink | [0...49] |
| 0.6 | 0021 | 0 | Format1 | Downlink | [0...49] |
| 0.7 | 0021 | 0 | Format1 | Downlink | [0...49] |
| 0.8 | 0021 | 0 | Format1 | Downlink | [0...49] |
| 0.9 | 0021 | 0 | Format1 | Downlink | [0...49] |

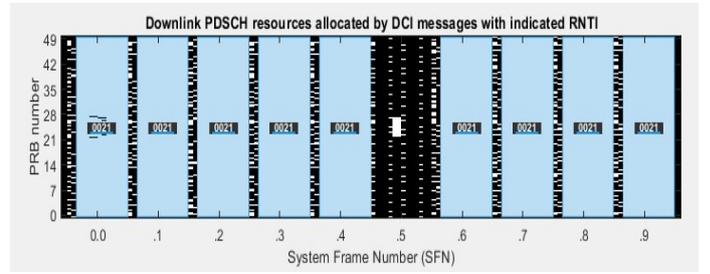

Fig. 5. RB allocation in PDSCHof signal without noise

## B. Simulation with random noise

Now add random noises to the signal for PDCCH and PDSCH with internal MATLAB LTE Toolbox module - OCNG (OFDMA Channel Noise Generator) Also will consider the non-zero frame number and more complex C-RNTI = 61, in LTE the numbers of C-RNTI are transformed into the hexadecimal number system.

Providing UE detection of new signal, obtained results shown in Figure 7. The resource block allocation is depicted at the Figure 6.

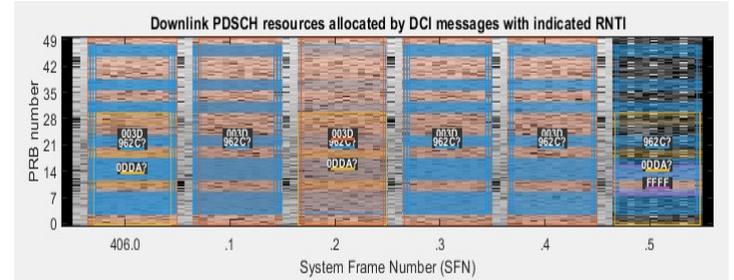

Fig. 6. RB allocation in PDSCH of signal with noise

| SFN | RNTI | NumErrors | DCIFormat | LinkDirection | PRBSet | Power |
|---|---|---|---|---|---|---|
| 406 | "003D" | 0 | "Format2" | "Downlink" | "[0...49]" | 20.99 |
| 406 | "962C" | 2 | "Format2" | "Downlink" | "[3...8 12...17 21...23 30...32 36...38 45...47]" | 20.43 |
| 406 | "962C" | 2 | "Format2" | "Downlink" | "[3...8 12...17 21...23 30...32 36...38 45...47]" | 20.43 |
| 406 | "962C" | 2 | "Format2" | "Downlink" | "[3...8 12...17 21...23 30...32 36...38 45...47]" | 20.43 |
| 406 | "0DDA" | 2 | "Format2D" | "Downlink" | "[0 10 11 18 19 27 29]" | 18.23 |
| 406 | "962C" | 2 | "Format2" | "Downlink" | "[3...8 12...17 21...23 30...32 36...38 45...47]" | 20.43 |
| 406.1 | "962C" | 2 | "Format2" | "Downlink" | "[3...8 12...17 21...23 30...32 36...38 45...47]" | 20.17 |
| 406.1 | "962C" | 2 | "Format2" | "Downlink" | "[3...8 12...17 21...23 30...32 36...38 45...47]" | 20.17 |
| 406.1 | "962C" | 2 | "Format2" | "Downlink" | "[3...8 12...17 21...23 30...32 36...38 45...47]" | 20.17 |
| 406.1 | "962C" | 2 | "Format2" | "Downlink" | "[3...8 12...17 21...23 30...32 36...38 45...47]" | 20.17 |
| 406.1 | "003D" | 0 | "Format2" | "Downlink" | "[0...49]" | 20.84 |
| 406.1 | "962C" | 2 | "Format2" | "Downlink" | "[3...8 12...17 21...23 30...32 36...38 45...47]" | 20.17 |
| 406.2 | "0DDA" | 2 | "Format2D" | "Downlink" | "[0 10 11 18 19 27 29]" | 18.36 |
| 406.2 | "962C" | 2 | "Format2" | "Downlink" | "[3...8 12...17 21...23 30...32 36...38 45...47]" | 20.35 |
| 406.2 | "003D" | 0 | "Format2" | "Downlink" | "[0...49]" | 20.92 |
| 406.3 | "962C" | 2 | "Format2" | "Downlink" | "[3...8 12...17 21...23 30...32 36...38 45...47]" | 20.31 |
| 406.3 | "003D" | 0 | "Format2" | "Downlink" | "[0...49]" | 20.93 |
| 406.3 | "962C" | 2 | "Format2" | "Downlink" | "[3...8 12...17 21...23 30...32 36...38 45...47]" | 20.31 |
| 406.3 | "962C" | 2 | "Format2" | "Downlink" | "[3...8 12...17 21...23 30...32 36...38 45...47]" | 20.31 |
| 406.3 | "962C" | 2 | "Format2" | "Downlink" | "[3...8 12...17 21...23 30...32 36...38 45...47]" | 20.31 |
| 406.4 | "003D" | 0 | "Format2" | "Downlink" | "[0...49]" | 20.9 |
| 406.4 | "962C" | 2 | "Format2" | "Downlink" | "[3...8 12...17 21...23 30...32 36...38 45...47]" | 20.21 |
| 406.4 | "962C" | 2 | "Format2" | "Downlink" | "[3...8 12...17 21...23 30...32 36...38 45...47]" | 20.21 |
| 406.4 | "962C" | 2 | "Format2" | "Downlink" | "[3...8 12...17 21...23 30...32 36...38 45...47]" | 20.21 |
| 406.5 | "FFFF" | 0 | "Format1A" | "Downlink" | "[8...15]" | 16.44 |
| 406.5 | "0DDA" | 2 | "Format2D" | "Downlink" | "[0 10 11 18 19 27 29]" | 11.62 |
| 406.5 | "962C" | 2 | "Format2" | "Downlink" | "[3...8 12...17 21...23 30...32 36...38 45...47]" | 10.67 |
| 406.5 | "962C" | 2 | "Format2" | "Downlink" | "[3...8 12...17 21...23 30...32 36...38 45...47]" | 10.67 |

Fig. 7. Decoded DCI of signal with noise

After adding noise, C-RNTI appeared which were decoded with errors. The script parameters set that the maximum number of non-matching bits is two, which means that all identifiers that were compared with the original DCI found more than two different bits are automatically considered decoding errors and are not included in the UE detection results. Also in this signal, there is "FFFF" RNTI. It is a system identifier and it is not used to identify subscriber devices. Knowing that the RNTI decoded without errors are correct, it is possible to exclude others, that have resource block allocation intersections with them. From here, the only user assigned resource blocks in this frame is a UE with C-RNTI 003D (hex) which corresponds to the parameter 61 in decimal numeric system

## IV. UE DETECTION FROM THE REAL BS SIGNAL

To capture signal form the BS software-defined radio platform NI USRP 2920 was used. Measuring will provided according to the algorithm, depicted at Figure 1. The spectrum of the signal is represented at Figure 8.

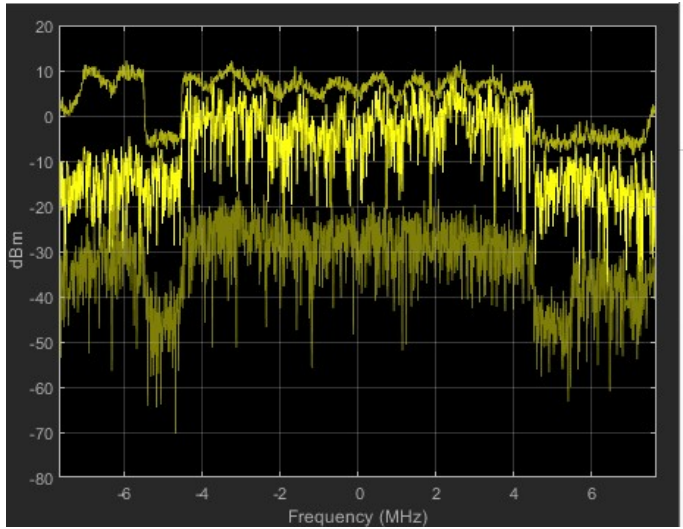

Fig. 8. Spectrum of the signal

Then decoding the MIB obtain that cell-wide settings:
- NDLRB: 50
- DuplexMode: 'FDD'
- CyclicPrefix: 'Normal'
- NCellID: 13
- NSubframe: 0
- CellRefP: 2
- PHICHDuration: 'Normal'
- NFrame: 867

Now the whole bandwidth are known, it is equal 50 RBs.

After, provide the PDCCH search and DCI decoding. Allocation of RBs is depicted at Figure 9.

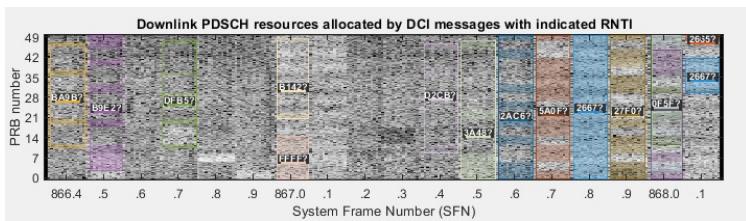

Fig. 9. RB allocation in PDSCH of real BS signal

The information from the decoded DCI message is shown in the table 3. Here there is one DCI with Format 0, which consist information about Uplink scheduling. In addition, the C-RNTI 2667 is repeated twice (for Downlink and Uplink) and occupy almost all RB in the subframes 2 and 8. There are also system RNTI "FFFF" in this frame. There are no any intersections of the resource block distribution; hence, it can be judged that the decoding was performed correctly.

TABLE III. DECODED DCI MESSAGE OF REAL BS

| SFN | RNTI | Num of errors | DCI Format | Link Direction | RB Set |
|---|---|---|---|---|---|
| 867.0 | FFFF | 1 | Format1A | Downlink | [0...14] |
| 867.0 | B142 | 1 | Format 2D | Downlink | [21 22 32 48] |
| 867.2 | 2667 | 2 | Format 0 | Uplink | [2...41] |
| 867.4 | D2CB | 2 | Format 2 | Downlink | [9 11 19 28 36 38 45 46] |
| 867.5 | 3A48 | 2 | Format 2C | Downlink | [0...8 21...23 45...47] |
| 867.6 | 2AC6 | 2 | Format 2D | Downlink | [0...5 12...14 18...20 24...26 30...32 42...44 48 49] |
| 867.7 | 5A0F | 2 | Format 2D | Downlink | [0...2 6...11 15...23 27...41 48 49] |
| 867.8 | 2667 | 2 | Format 2 | Downlink | [0...49] |
| 867.9 | 27F0 | 1 | Format 2C | Downlink | [0...5 9...14 18...20 30...35 39...44 48 49] |

There are 7 unique identifiers in this frame, it means that there are 7 active UEs at this moment. There are 154 RBs from 500 it total which are used in this frame. Information load is equal:

$$I_{frame} = 154 / 500 * 100\% = 30.8\% \qquad (2)$$

It means that the channel used only for 1/3 and there are reserve resource blocks if the load will increase in the maximum busy hour.

## V. Conclusions

In the course of the work, the method of determining the information load of the LTE network based on decoding DCI messages in PDCCH was studied and tested. This approach has advantages over other methods of channel estimation, as it does not require direct access to the mobile operator's equipment and allows real-time measurements. The algorithm is implemented based on the USRP software-defined radio platform and a personal computer with MATLAB software. Also, the recording of the signal from a real LTE base station was analyzed and the results of DCI decoding and evaluation of the information load of one frame of this signal were obtained. At this stage of work, the distribution per frame can be estimated, but for an objective assessment of the load of hundreds, it needs to analyze statistics for a long period (no less than 10 minutes), as well as at different times of the day, including in the busy hour. This task will form the basis for further research. Then a model of the cell as a queuing system will be built. Restoring the average number of unique requests, their size, the intensity of arrival and service. Based on these parameters, we will be able to get averaged knowledge on the probability of occupying a certain number of channels, the probability of blocking, the average throughput of the cell, and other parameters that determine the quality of service.

## Acknowledgment

This project is supported by the Ministry of Science and Higher Education of the Russian Federation, agreement № 075-03-2020-051 (topic № fzsu-2020-0020).